# Ni(OH)$_2$ Nanoplates Grown on Graphene as Advanced Electrochemical Pseudocapacitor Materials

Hailiang Wang, Hernan Sanchez Casalongue, Yongye Liang and Hongjie Dai[*]

*Department of Chemistry and Laboratory for Advanced Materials, Stanford University, Stanford, CA 94305, USA*

*[*] Correspondence to hdai@stanford.edu*

**Abstract**

Ni(OH)$_2$ nanocrystals grown on graphene sheets with various degrees of oxidation are investigated as electrochemical pseudocapacitor materials for potential energy storage applications. Single-crystalline Ni(OH)$_2$ hexagonal nanoplates directly grown on lightly-oxidized, electrically-conducting graphene sheets (GS) exhibit a high specific capacitance of ~1335F/g at a charge and discharge current density of 2.8A/g and ~953F/g at 45.7A/g with excellent cycling ability. The high specific capacitance and remarkable rate capability are promising for applications in supercapacitors with both high energy and power densities. Simple physical mixture of pre-synthesized Ni(OH)$_2$ nanoplates and graphene sheets show lower specific capacitance, highlighting the importance of direct growth of nanomaterials on graphene to impart intimate interactions and efficient charge transport between the active nanomaterials and the conducting graphene network. Single-crystalline Ni(OH)$_2$ nanoplates directly grown on graphene sheets also significantly outperform small Ni(OH)$_2$ nanoparticles grown on heavily-oxidized, electrically-insulating graphite oxide (GO), suggesting that the electrochemical performance of these composites are dependent on the quality of graphene substrates and the morphology and crystallinity of the nanomaterials grown on top. These results suggest the importance of rational design and synthesis of graphene-based nanocomposite materials for high-performance energy applications.

**Introduction**

The increasing demand for energy and growing concerns about air pollution and global warming have stimulated intense research on energy storage and conversion from alternative energy sources.[1-3] Supercapacitors are considered as a promising candidate for energy storage due to high power performance, long cycle life and low maintenance cost.[4-6] While supercapacitors are ideal for applications that require short-term power boost, such as emergency power supply and peak power assistance for batteries in electric vehicles, it is highly desirable to increase the energy density of supercapacitors to approach that of batteries, which could enable their use as primary power sources. Pseudo-capacitive materials such as hydroxides,[7-10] oxides[11-20] and polymers[21-23] are being explored for producing supercapacitors with increased specific capacitance and high energy density. However, such 'pseudocapacitors' often result in compromises of rate capability and reversibility because they rely on faradic redox reactions and the active materials are typically too insulating to support fast electron transport required by high rates.

Graphene is a two-dimensional material with high surface area and electrical conductivity, light weight, high flexibility and mechanical strength. Graphene is an ideal single-atom-thick substrate for growth of functional nanomaterials[23-26] to render them electrochemically active and electrically conducting to the outside current collectors. Recent work have shown Li-ion battery and supercapacitor applications of oxides[24,25] and polymers[23] coupled with reduced graphite oxide. However, Graphite oxide (GO) remains highly resistive even after reduction, which is not optimal for energy storage applications. It remains highly interesting to boost the performance of graphene based energy storage by growing nanomaterials with well defined morphology and crystallinity on highly pristine and electrically conducting graphene.

Recently we reported a two-step method to grow $Ni(OH)_2$ nanocrystals on graphene with various degrees of oxidation including lightly oxidized, highly conducting graphene sheets (GS).[26] The morphology, size and crystallinity of the nanocrystals can be tuned by the surface chemistry of the underlying graphene substrates.[26] $Ni(OH)_2$ has been a primary electrode material in alkaline batteries. It is also an attractive candidate in supercapacitor applications[8-10] due to high theoretical specific capacitance, well defined redox behavior, and low cost. Here, we show $Ni(OH)_2$ hexagonal nanoplates grown on GS with low oxidation for potential supercapacitor applications with both high power and energy capabilities. In the graphene composite material, single-crystalline thin nanoplates of $Ni(OH)_2$ are selectively and directly grown on highly conducting graphene. The graphene sheets overlap with each other to afford a three-dimensional conducting network for fast electron transfer between the active materials and the charge collector. We found that the degree of oxidation of the graphene substrates is important to the pseudo-capacitance and rate capability of the composite materials. The morphology and crystallinity of the $Ni(OH)_2$ nanocrystals play important roles as well. Highly crystalline $Ni(OH)_2$ hexagonal nanoplates grown on low-oxidation GS showed the best electrochemical characteristics as pseudocapacitor materials with a specific capacitance of ~952F/g even at a high charge/discharge rate of 45.7A/g. The energy density was estimated to be ~37Wh/kg at a power density of ~10kW/kg in a voltage range of 0.55V. In contrast, $Ni(OH)_2$ nanocrystals grown on highly-oxidized GO and $Ni(OH)_2$ nanoplates physically mixed with graphene sheets both showed inferior performance than $Ni(OH)_2$ nanoplates directly synthesized on highly conducting graphene sheets.

**Results**

The $Ni(OH)_2$/graphene composites were synthesized via a two-step method (see Supporting Information for details), as reported recently.[26] Uniform coatings of small $Ni(OH)_2 \cdot 0.75H_2O$ nanoparticles were first deposited on GS (or GO) by hydrolysis of $Ni(CH_3COO)_2$ at 80℃ in a 10:1 $N$, $N$- dimethylformamide (DMF)/$H_2O$ mixed solvent. The intermediate products were collected and dispersed in pure $H_2O$ for hydrothermal treatment at 180℃. The small particle coatings on GS were found to diffuse and recrystallize into well-defined hexagonal nanoplates of β-$Ni(OH)_2$ (Figure 1a, 1b, S1a) during the second hydrothermal step. The side length of the $Ni(OH)_2$ nanoplates was several hundred nanometers (Figure 1a, 1b) and the thickness was several nanometers.[26] In contrast, due to strong interactions with the dense functional groups and defects on GO surface, the $Ni(OH)_2 \cdot 0.75H_2O$ precursor coating on GO transformed into small nanoparticles of β-$Ni(OH)_2$ (Figure 1c, 1d, S1b) during hydrothermal treatment, without forming large single-crystalline nanoplates as in the GS case.[26] These small nanoparticles of β-$Ni(OH)_2$ were less crystalline than the nanoplates formed on GS, revealed by transmission electron microscopy (TEM) and X-ray diffraction (XRD) (Figure S1a,

S1b).[26] Both the Ni(OH)$_2$/GS and Ni(OH)$_2$/GO composites contained ~30% of graphene by mass (see Supporting Information for details). We have also synthesized Ni(OH)$_2$ hexagonal nanoplates without any graphene added (Figure 1e) by the same two-step method (see Supporting Information for details). These nanoplates grown in free solution exhibited similar morphology and crystallinity as the nanoplates grown on GS (Figure 1e,S1c).

About ~1mg of Ni(OH)$_2$/graphene composites were deposited and compressed into a Ni foam support (without using any carbon black additives) for electrochemical measurements (see Supporting Information for details) in a three-electrode beaker cell with a Ag/AgCl (in 3M NaCl) reference electrode and 1M KOH aqueous electrolyte. Figure 2a shows Cyclic voltammetry (CV) curves of the Ni(OH)$_2$/GS composite at various scan rates. The redox current peaks corresponded to the reversible reactions of Ni(II) $\leftrightarrow$ Ni(III) (Figure 2a).[8-10] Note that background signal due to the Ni foam was negligible (Figure S3). The average specific capacitance of the Ni(OH)$_2$ nanoplates on GS was calculated to be ~1267 F/g (based on mass of Ni(OH)$_2$, ~887 F/g based on total sample mass) at a scan rate of 5mV/s (Figure 2b), and ~877F/g at a high scan rate of 40mV/s, ~70% of that at 5mV/s. Figure 2c shows galvanostatic discharge curves of the Ni(OH)$_2$/GS composite at various current densities. The hexagonal Ni(OH)$_2$ nanoplates grown on GS showed a specific capacitance as high as ~1335F/g (based on mass of Ni(OH)$_2$, ~935F/g based on total sample mass) at a charge and discharge current density of 2.8A/g (Figure 2c, 2d). The specific capacitance was still as high as ~953F/g even at a high charge and discharge current density of 45.7A/g (Figure 2c, 2d). Importantly, the Columbic efficiency was nearly 100% for each cycle of charge and discharge (Figure 2e).

There was no obvious capacitance decrease observed over 2000 cycles of charge and discharge at a high current density of 28.6A/g (Figure 2f). These results revealed the high specific capacitance and remarkable rate capability of the $Ni(OH)_2$/GS composite material for high-performance electrochemical pseudo-capacitors.

We carried out similar electrochemical measurements for $Ni(OH)_2$ hexagonal nanoplates grown in free solution and then physically mixed with GS in the same Ni/C ratio as the synthesized $Ni(OH)_2$/graphene composite material (Figure 1e, 1f). The simple physical mixture of GS and $Ni(OH)_2$ exhibited lower specific capacitance than $Ni(OH)_2$ nanoplates grown on GS (Figure 3). The average specific capacitance was ~484F/g (based on mass of $Ni(OH)_2$, ~339F/g based on total sample mass) at a scan rate of 40mV/s (Figure 3b), a factor of ~1.8 lower than that of the $Ni(OH)_2$ nanoplates grown on GS at the same scan rate.

Also in strong contrast to the $Ni(OH)_2$ hexagonal nanoplates grown on GS, $Ni(OH)_2$ nanoparticles grown on GO showed much lower specific capacitance and inferior rate capability, as revealed by CV and galvanostatic measurement (Figure 4). The average specific capacitance was ~425 F/g (based on mass of $Ni(OH)_2$, ~297 F/g based on total sample mass) at a scan rate of 5mV/s (Figure 4b). At a high scan rate of 40mV/s, the average specific capacitance was only ~255F/g, ~60% of that at 5mV/s. At a charge and discharge current density of 1.4A/g, the $Ni(OH)_2$ nanoparticles grown on GO showed a specific capacitance of ~445F/g (based on mass of $Ni(OH)_2$, ~312F/g based on total sample mass) (Figure 4c, 4d). The specific capacitance further decreased to ~263F/g at a current density of 14.3A/g (Figure 4c, 4d).

Figure 5 shows the Ragone plot (power density vs. energy density) of the $Ni(OH)_2$ nanocrystals grown on GS and GO, and pre-synthesized $Ni(OH)_2$ nanoplates mixed with GS and GO (with reference to the Ag/AgCl reference electrode).[20] The energy and power densities were derived from CV curves at various scan rates (see Supporting Information for details). Our single-crystalline $Ni(OH)_2$ hexagonal nanoplates grown on GS delivered a high energy density of ~37Wh/kg at a high power density of ~10kW/kg, superior to $Ni(OH)_2$ nanoplates simply mixed with GS and $Ni(OH)_2$ nanocrystals grown on GO (Figure 5). Note that these characteristics corresponded to a narrow potential window of 0.55V against the Ag/AgCl reference electrode. It is highly desirable to couple our $Ni(OH)_2$/GS material with a suitable counter electrode material with comparably high performance to achieve a large operating voltage range and optimize the energy and power densities of real supercapacitors. This is clearly the next step for the $Ni(OH)_2$/graphene materials.

**Discussion**

Our single-crystalline $Ni(OH)_2$ hexagonal nanoplates grown on highly conducting GS exhibited excellent electrochemical characteristics and high cycling stability, making them potentially useful for high performance supercapacitor materials. Measured by the same CV or galvanostatic method in three-electrode systems, our material showed higher stable specific capacitance at higher charge/discharge rates than many of the previously reported pseudo-capacitive nanomaterials including $Ni(OH)_2$,[8-10] $NiO$,[11] $MnO_2$,[13] $Mn_3O_4$,[16] $RuO_2$[18] and their composites with carbon nanotubes[12,14,15,17,19,20]. The highest specific capacitance measured for $Ni(OH)_2$ was ~1600F/g at 4A/g current density after 300 cycles (with a slowly decreasing trend)[10]. At higher current densities, our materials exhibited higher specific capacitance than in ref. 10. The characteristics of high specific capacitances, high energy and power densities, and high cycling stabilities are all critical to high performance electrochemical supercapacitors.

Several features make the single-crystalline $Ni(OH)_2$ hexagonal nanoplates grown on GS unique building blocks for high-capacity and ultrafast energy storage and releasing. First, the $Ni(OH)_2$ nanoplates are directly grown and anchored on graphene sheets. The interactions between $Ni(OH)_2$ nanoplates and GS could be both covalent chemical bonding and van der Waals interactions, at oxygen-containing defect sites and pristine regions of the GS respectively. This intimate binding affords facile electron transport between individual nanoplates and the GS [by themselves, $Ni(OH)_2$ nanocrystals are electrically insulating], which is key to both high specific capacitance and rate capability of the $Ni(OH)_2$/GS material. With little 'dead' volume, most of the $Ni(OH)_2$ nanoplates in the macroscopic ensemble are electrochemically active through the graphene network. Rapid charge transport from the nanoplates to the underlying graphene affords fast redox reactions at high scan rates and charge and discharge currents. For the physical mixture of pre-synthesized $Ni(OH)_2$ nanoplates and GS, the inferior electrochemical characteristics were mainly due to much less intimate contact between the $Ni(OH)_2$ nanoplates and GS. Phase separation was in fact observed between the $Ni(OH)_2$ nanoplates and GS in the mixture (Figure 1f).

We used high-quality graphene sheets[27-29] as growth substrates for $Ni(OH)_2$ nanoplates. Graphene sheets with low oxidation are needed to imparting excellent electrical conductivity to the macroscopic ensemble of the composite materials (without the need of carbon black additives). The highly conducting GS network allows rapid and effective charge transport between the $Ni(OH)_2$ nanoplates in the macroscopic ensemble and the current collector, allowing for fast energy storage and releasing. Unlike graphene sheets, GO is poorly conducting.[27-31] For macroscopic electrodes made of packed $Ni(OH)_2$/GO, electron transport through the GO network is highly resistive. As a result, $Ni(OH)_2$/GO composites show significantly lower specific capacitance than $Ni(OH)_2$/GS composites at all scan rates (Figure 4). Graphite oxide used to make the $Ni(OH)_2$/GO composite contains ~20% of oxygen, much more than that in GS (~5%).[26,27,29] Even after reduction, the GO was still ~50 times less conducting than the GS.[26-29] The high concentration of functional groups and defects, as well as the poorly-conducting nature of the GO hinders fast and efficient electron transport from the active $Ni(OH)_2$ to the charge collector via the GO network, which results in low specific capacitance and poor rate capability. To confirm this scenario, we measured the conductivity of films of $Ni(OH)_2$/GS and $Ni(OH)_2$/GO composites by two probe electrical measurements. The latter was found to be ~50 times more resistive than the former, confirming the highly resistive nature of the $Ni(OH)_2$/GO composite material.

In another control experiment, we mixed pre-synthesized $Ni(OH)_2$ hexagonal nanoplates with GO and GS respectively at the same Ni/C ratio (see Supporting Information for details), and compared the electrochemical performance of these mixtures. The specific capacitance of $Ni(OH)_2$/GO mixture was a factor of ~1.6 lower than that of the $Ni(OH)_2$/GS mixture at the same scan rate of 40mV/s (Figure 3, S2). This further confirmed the importance of graphene quality and conductivity to the pseudo-capacitive performance of $Ni(OH)_2$.

The large, thin morphology (<10nm) and single-crystalline nature of $Ni(OH)_2$ nanoplates grown on GS is ideal to afford effective charging and discharging through the nanoplate thickness during fast cycling.[4] We observed that pre-synthesized $Ni(OH)_2$ hexagonal nanoplates mixed with GO showed higher specific capacitance (~1.2 times at a scan rate of 40mV/s) than $Ni(OH)_2$ nanoparticles directly grown on GO (Figure S2, 4). This suggested that morphology and crystallinity were contributing factors to the different electrochemical performances exhibited by the $Ni(OH)_2$/GS and $Ni(OH)_2$/GO composites. Single-crystalline thin nanoplates showed more favorable electrochemical characteristics over the less crystalline small nanoparticles with high specific capacitance. Taken together, we conclude that it is highly important to use high-quality/conductivity graphene as growth substrate and control the morphology and crystallinity of the nanomaterials grown on graphene to produce advanced graphene/nanocrystal composite materials for energy applications.

**Conclusion**

We demonstrated $Ni(OH)_2$/GS composite as an interesting material for electrochemical pseudocapacitors with potentially high energy density, high power density and long cycle life. single-crystalline $Ni(OH)_2$ hexagonal nanoplates grown on GS showed high specific capacitance and remarkable rate capability, significantly outperforming $Ni(OH)_2$ nanoparticles grown on GO and $Ni(OH)_2$ nanoplates simply mixed with GS. We also showed that the quality of graphene and the morphology and crystallinity of the nanomaterials are both important to the high electrochemical performance of these graphene based composite materials for energy storage. It is highly desirable to couple our $Ni(OH)_2$/GS material with a suitable counter electrode material with comparably high performance to achieve a large operating voltage range and optimize the energy and power densities of real supercapacitors. This will be the next step for the $Ni(OH)_2$/graphene materials.

**Methods**

The $Ni(OH)_2$/GS (GO) composites were made by a previously reported two-step method.[26] In a typical synthesis, 20ml of GS (GO)/DMF suspension with a concentration of ~0.25mg/ml graphene was heated to 80℃, to which 2ml of $Ni(Ac)_2$ aqueous solution with a concentration of 0.2M was added. The suspension was kept at 80℃ with stirring for 1 hour. The intermediate products were transferred to 20ml of water and sealed in Teflon lined stainless steel autoclaves for hydrothermal reaction at 180℃ for 10 hours. The free $Ni(OH)_2$ nanoplates were made by the same two-step method without any graphene added. The synthesized materials were characterized by SEM (FEI XL30 Sirion), TEM (FEI Tecnai F20), and XRD (PANalytical X'Pert). For electrochemical measurements, ~1mg of the materials were dispersed in ethanol with 1% PTFE binder (without any other carbon additive), and then drop-dried into a Ni foam. The Ni foam was baked at 80℃ and compressed. CV and galvanostatic charge and discharge were carried out on a CHI 660D electrochemistry workstation. The specific capacitances were calculated from CV and galvanostatic curves. The energy and power densities were derived from CV curves. Detailed experimental, data analysis and supplementary figures are available in Supporting Information.

**Acknowledgements**

This work was supported in part by MARCO MSD Focus Center, Intel, Office of Naval Research and an NSF instrumentation grant to Stanford Chemistry. We thank Professor Christopher Chidsey for use of equipment and helpful discussion.

**Figures**

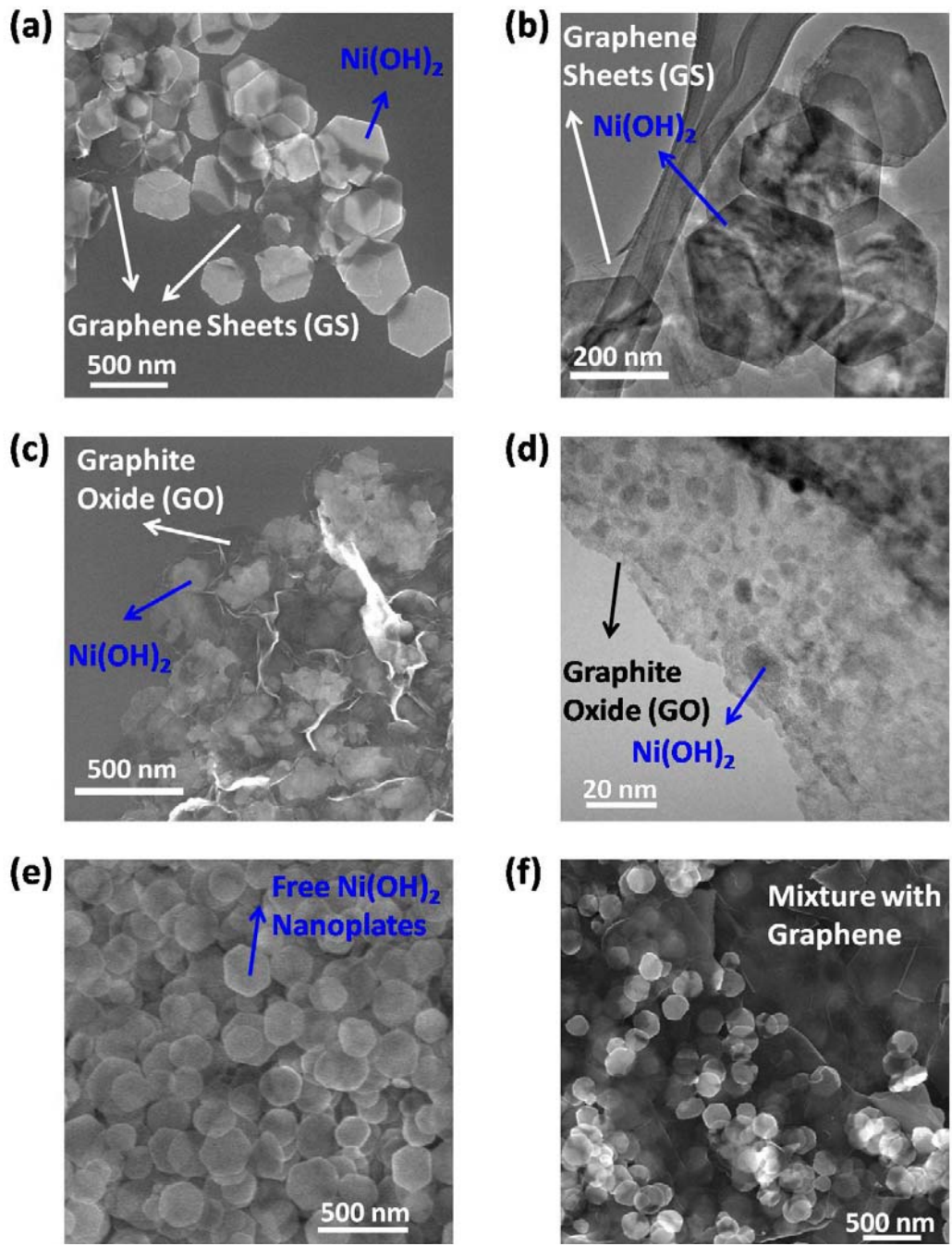

**Figure 1.** SEM and TEM characterizations of Ni(OH)$_2$/GS composite, Ni(OH)$_2$/GO composite, and Ni(OH)$_2$ + GS physical mixture. (a) SEM image of Ni(OH)$_2$ nanoplates grown on GS. (b) TEM image of Ni(OH)$_2$ nanoplates grown on GS. (c) SEM image of Ni(OH)$_2$ nanoparticles grown on GO. (d) TEM image of Ni(OH)$_2$ nanoparticles grown on GO. (e) SEM image of Ni(OH)$_2$ hexagonal nanoplates grown in free solution (without graphene). (f) SEM images of simple physical mixture of pre-synthesized free Ni(OH)$_2$ nanoplates and GS.

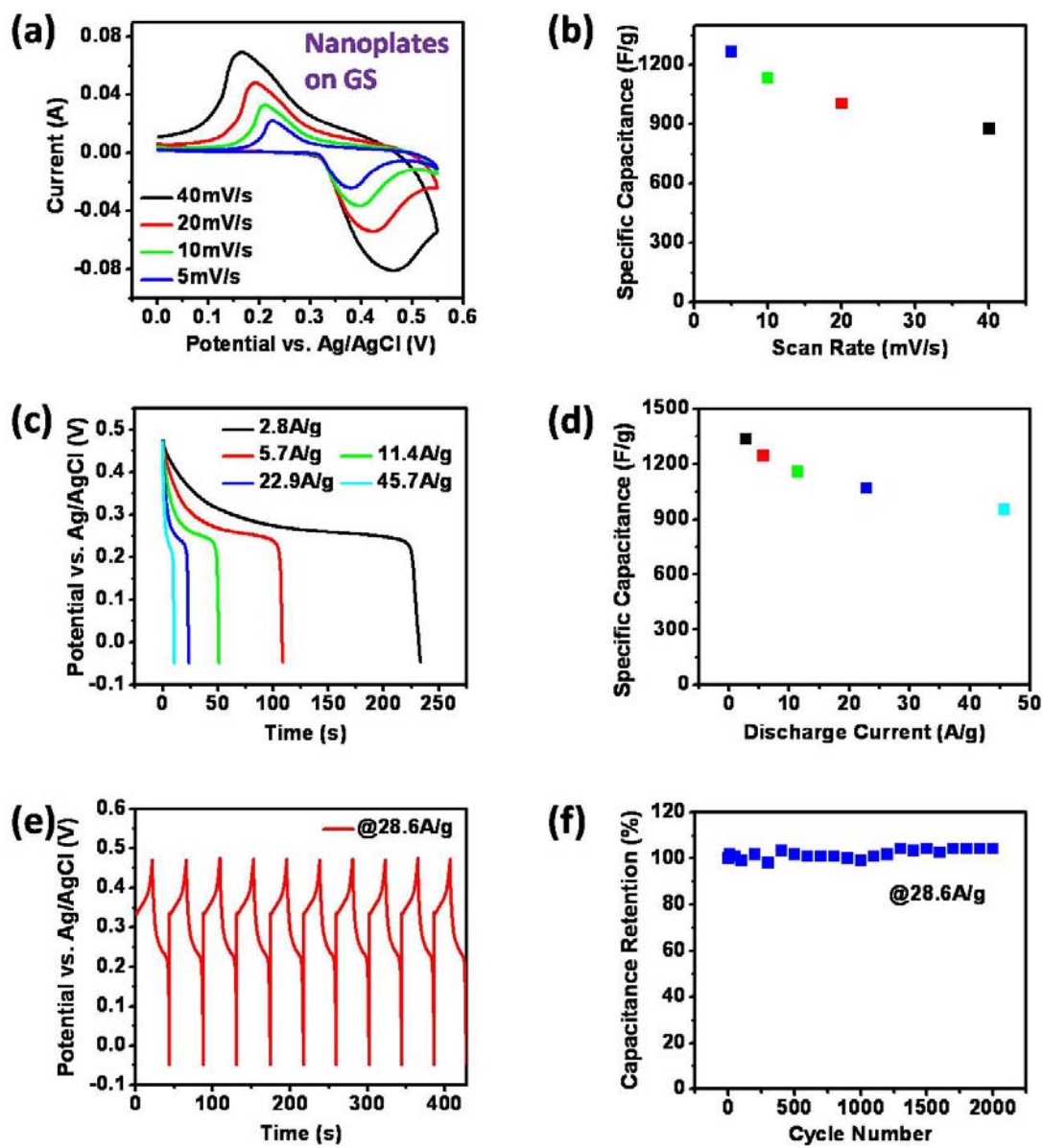

**Figure 2.** Electrochemical characterizations of Ni(OH)$_2$ hexagonal nanoplates grown on GS. (a) CV curves of Ni(OH)$_2$/GS composite at various scan rates. (b) Average specific capacitance of Ni(OH)$_2$ nanoplates grown on GS (~1mg combined mass) at various scan rates. (c) Galvanostatic discharge curves of Ni(OH)$_2$ nanoplates grown on GS at various discharge current densities. (d) Average specific capacitance of Ni(OH)$_2$ nanoplates grown on GS at various discharge current densities. (e) Galvanostatic charge and discharge curves of Ni(OH)$_2$ nanoplates grown on GS at a current density of 28.6A/g. (f) Average specific capacitance versus cycle number of Ni(OH)$_2$/GS at a galvanostatic charge and discharge current density of 28.6A/g.

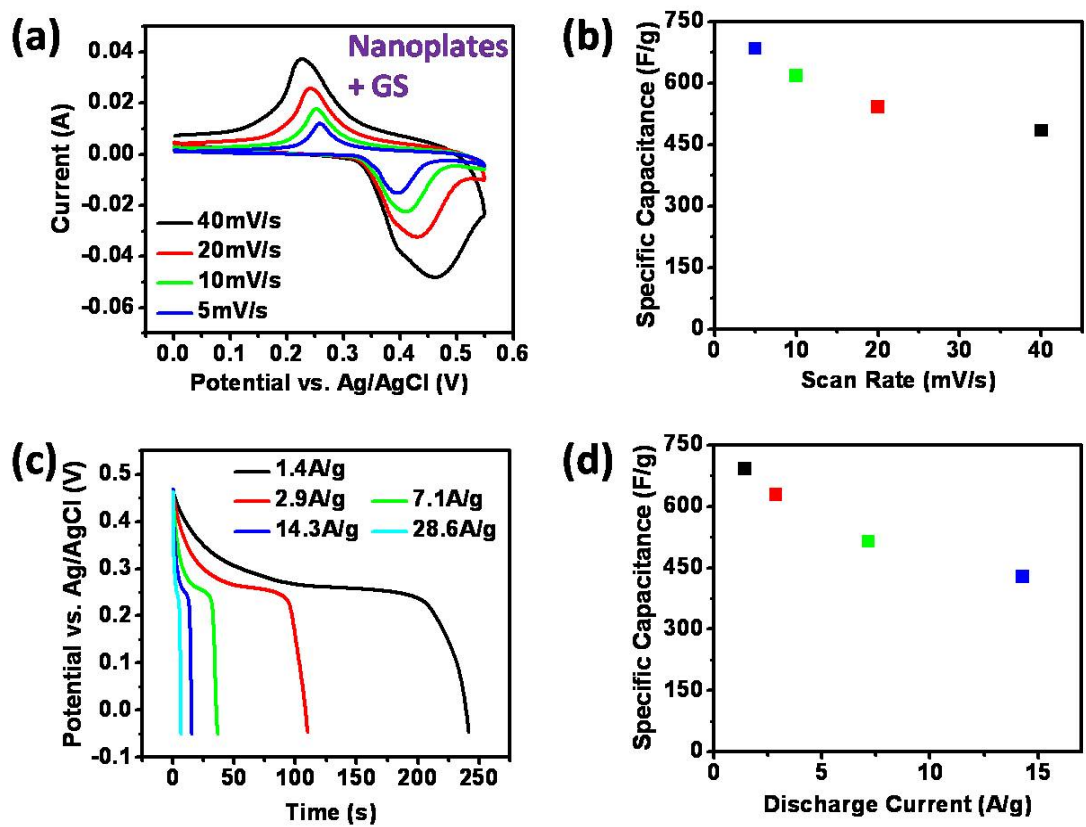

**Figure 3.** Electrochemical characterizations of simple physical mixture of Ni(OH)$_2$ nanoplates and GS. (a) CV curves of Ni(OH)$_2$ GS mixture at various scan rates. (b) Average specific capacitance of Ni(OH)$_2$ GS mixture (~1mg combined mass) at various scan rates. (c) Galvanostatic discharge curves of Ni(OH)$_2$ GS mixture at various discharge current densities. (d) Average specific capacitance of Ni(OH)$_2$ GS mixture at various discharge current densities.

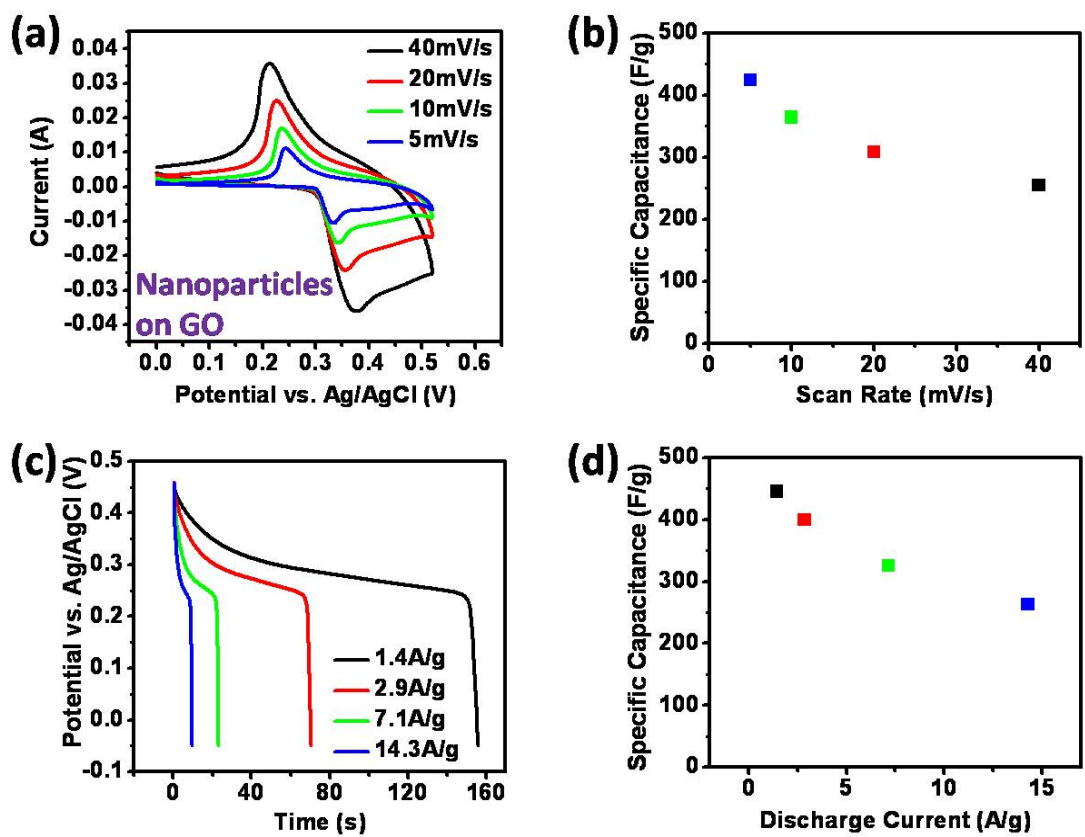

**Figure 4.** Electrochemical characterizations of Ni(OH)$_2$ nanoparticles grown on GO. (a) CV curves of Ni(OH)$_2$/GO composite at various scan rates. (b) Average specific capacitance of Ni(OH)$_2$ nanoparticles grown on GO (~1.5mg combined mass) at various scan rates. (c) Galvanostatic discharge curves of Ni(OH)$_2$ nanoparticles grown on GO at various discharge current densities. (d) Average specific capacitance of Ni(OH)$_2$ nanoparticles grown on GO at various discharge current densities.

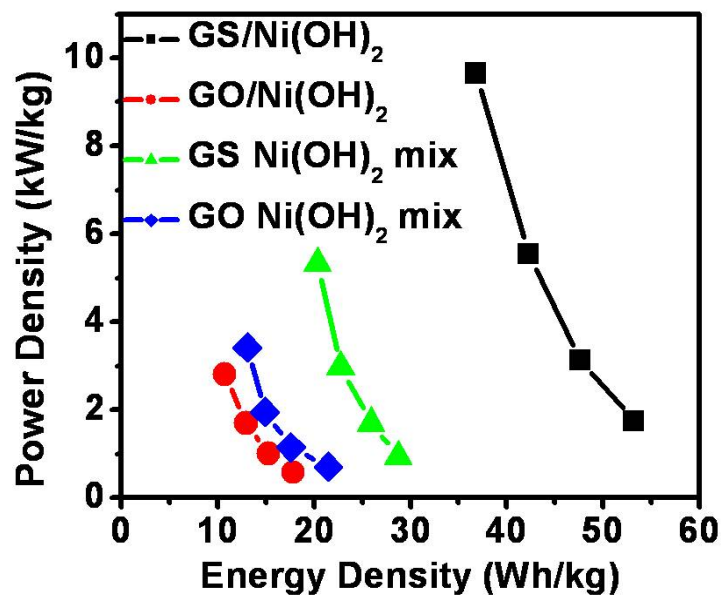

**Figure 5.** Ragone plot (power density vs. energy density) of $Ni(OH)_2$ hexagonal nanoplates grown on GS (black), $Ni(OH)_2$ nanoparticles grown on GO (red), and pre-synthesized $Ni(OH)_2$ hexagonal nanoplates physically mixed GS (green) and GO (blue). The energy and power densities were derived from the CV curves at variously scan rates (see Supporting Information for details).